\documentstyle[12pt]{article}
\begin{document}

\title{\Large\bf Derivative expansion and large gauge invariance at
finite temperature} 

\author{J. Barcelos-Neto\thanks{\noindent e-mail: 
barcelos@if.ufrj.br}\\ 
Instituto de F\'{\i}sica\\ 
Universidade Federal do Rio de Janeiro\\ 
RJ 21945-970 - Caixa Postal 68528 - Brasil\\
\\
Ashok Das\\
Department of Physics and Astronomy\\
University of Rochester\\
Rochester, NY 14627 - USA
\\}
\date{}

\maketitle
\abstract
We study the $0+1$ dimensional Chern-Simons theory at finite
temperature within the framework of derivative expansion. We obtain
various interesting relations, solve the theory within this framework
and argue that the derivative expansion is not a suitable formalism
for a study of the question of large gauge invariance.

\vfill
\vspace{1cm}
\newpage

\section{Introduction}

\bigskip
The question of large gauge invariance in $2+1$ dimensional
Chern-Simons theories at finite temperature has been of interest for
some time now \cite{Deser1}--\cite{Bra}. Simply put, at finite
temperature, the coefficient of the Chern-Simons term becomes
temperature dependent \cite{Ishi,Babu} making it incompatible with
the quantization condition necessary for large gauge invariance to
hold. More recently, this question has been successfully analyzed
\cite{Du} in the context of the $0+1$ dimensional Chern-Simons theory
which suggests a possible solution for the puzzle of large gauge
invariance in $2+1$ dimensions. Of particular interest is the fact
that, at finite temperature, new terms seem to be generated in the
effective action which are higher order (in the field variables) and
are nonextensive in nature restoring the large gauge invariance in
the full effective action. This is really a very interesting scenario
and, consequently, this model has already been analyzed from other
points of view \cite{Das1} for a better understanding of various
features and, furthermore, the results of this analysis have already
been generalized to higher dimensions as well under various
assumptions \cite{Deser2}--\cite{Gon}. However, a complete and
satisfactory understanding of the question of large gauge invariance
in the $2+1$ dimensional theory is yet to come. And, for this reason,
it is worth analyzing the new features in the $0+1$ dimensional model
from as many points of view as is possible.

\medskip 
In this paper, we study this model within the framework of the
derivative expansion \cite{Ait2,Das2}. This is important from various
points of view.  First, one of the earliest calculations of the
temperature dependence of the Chern-Simons coefficient was, in fact,
in the context of the derivative expansion \cite{Babu}. Second, in
any higher dimensional theory, an exact solution is not possible.
Consequently, one has to resort to some approximation scheme and the
derivative expansion  has proved to be an excellent approximation
scheme in the study of several interesting phenomena. However, by
construction, the effective action in the context of derivative
expansion is an extensive quantity. And, so, the natural question
that arises is how one would see nonextensive contributions to the
effective action at finite temperature within this framework. Third,
the $0+1$ dimensional model is free from the problems of
nonanalyticity that plague the derivative expansion in higher
dimensions at finite temperature \cite{Das3}. In this paper, we solve
the $0+1$ dimensional model at finite temperature within the
framework of the derivative expansion. In some sense, this analysis
is complementary to the one in ref. \cite{Das1}. We derive various
interesting relations that will be useful in the study of finite
temperature field theory. Even though the $0+1$ dimensional model can
be solved within this framework, we argue that the derivative
expansion is not the appropriate framework for the study of large
gauge invariance in higher dimensions.

\vspace{1cm}
\section{Derivative expansion at T=0} 
\renewcommand{\theequation}{2.\arabic{equation}}
\setcounter{equation}{0}

\bigskip 
The $0+1$ dimensional theory is described by the Lagrangian

\begin{equation}
L=\bar\psi_j\,\bigl(i\partial_t-m-A\bigr)\,\psi_j-\kappa\,A
\label{2.1}
\end{equation}

\bigskip\noindent 
where the flavor index, $j=1,2,\dots,N_f$ is being summed. We follow
the notations of ref. \cite{Das1} to which we refer the readers for
details. The contribution of the fermions to the effective action is
given by

\begin{eqnarray}
\Gamma\,[A]&=&-\,i\,\ln\,\biggl[
\frac{\det\,\bigl(i\partial_t-m-A\bigr)}
{\det\,\bigl(i\partial_t-m\bigr)}\biggr]^{N_f}
\nonumber\\
&=&-\,i\,N_f\,\ln\,\det\,\Bigl[1+\,i\,S_F\,A\Bigr]
\label{2.2}
\end{eqnarray}

\bigskip\noindent
which is normalized to vanish when the external gauge field, $A$,
vanishes. Here, $S_F$ represents the fermion propagator.

\medskip
Using the relation

\begin{equation}
\det\,O=\exp\,{\rm Tr}\,\ln\,O
\label{2.3}
\end{equation}

\bigskip\noindent
we can write

\begin{eqnarray}
\Gamma\,[A]&=&-\,i\,N_f\,{\rm Tr}\,\ln\,\bigl[1+iS_F\,A\bigr]
\nonumber\\
&=&-\,i\,N_f\,{\rm Tr}\,\Bigl[iS_FA
-\frac{1}{2}\,iS_FA\,iS_FA+\dots\Bigr]
\label{2.4}
\end{eqnarray}

\bigskip\noindent 
Here ``Tr" simply stands for the trace in any complete basis (there
is no Dirac index in $0+1$ dimensions). In the derivative expansion,
$S_F(p)$ and $A$ are treated as noncommuting operators in (\ref{2.4})
which can be commuted past each other through the use of the standard
and canonical commutation relation

\begin{equation}
Ap=pA+i\,(\partial_tA)
\label{2.5}
\end{equation}

\bigskip
At zero temperature, the fermion propagator, in this model, is given 
by

\begin{equation}
S_F(p)=\frac{i}{p-m+i\epsilon}
\label{2.6}\\
\end{equation}

\bigskip\noindent 
so that the effective action at zero temperature (due to the
fermions) has the form

\begin{equation}
\Gamma\,[A]=-\,i\,N_f\,{\rm Tr}\,\biggl[
-\,\frac{1}{p-m+i\epsilon}\,A
-\frac{1}{2}\,\frac{1}{p-m+i\epsilon}\,A
\,\frac{1}{p-m+i\epsilon}\,A+\cdots\biggr]
\label{2.7}
\end{equation}

\bigskip\noindent
If we look at the first term on the right hand side, we obtain

\begin{eqnarray}
iN_f\,{\rm Tr}\,\frac{1}{p-m+i\epsilon}\,A
&=&iN_f\,\int\frac{dp}{2\pi}\,\frac{1}{p-m+i\epsilon}\,
\int dt\,A(t)
\nonumber\\
&=&\frac{N_f}{2}\int dt A(t)
\nonumber\\
&=&\frac{N_f}{2}\,a
\label{2.8}
\end{eqnarray}

\bigskip\noindent 
It is worth noting here that the integrand of the momentum integral
does not satisfy the Cauchy convergence condition which implies that
the correct value of the integral is obtained only after taking the
contribution of the semi-circle into account.

\medskip
The higher order terms in Eq. (\ref{2.7}) can be brought to a form,
through the use of Eq. (\ref{2.5}), where all the momentum factors
are at the left and, in general, will have the form

\begin{equation}
N_f\,{\rm Tr}\,\sum_{n=2}^\infty
\frac{1}{(p-m+i\epsilon)^n}\,f_n(A)
=N_f\sum_{n=2}^\infty\int\frac{dp}{2\pi}\,
\frac{1}{(p-m+i\epsilon)^n}\int dt\,f_n(A)
\label{2.9}
\end{equation}

\bigskip\noindent 
where $f_n(A)$ is a functional of $A$ involving derivatives as well.
It is clear from the structure  of the momentum integral in Eq.
(\ref{2.9}) that it vanishes. This is easily seen by closing the
contour in the upper half of the complex plane in which case, there
is no singularity inside the contour. Alternately,

\begin{eqnarray}
\int\frac{dp}{2\pi}\,\frac{1}{(p-m+i\epsilon)^n}
&=&\frac{1}{(n-1)!}\,\frac{\partial^{n-1}}{\partial m^{n-1}}
\int\frac{dp}{2\pi}\,\frac{1}{(p-m+i\epsilon)}
\nonumber\\
&=&\frac{1}{(n-1)!}\,\frac{\partial^{n-1}}{\partial m^{n-1}}\,
\Bigl(-\,\frac{i}{2}\Bigr)=0\,\,\,\,{\rm for}\,\,n\geq2
\label{2.10}
\end{eqnarray}

\bigskip\noindent
This shows that all the higher order terms in Eq. (\ref{2.7}) which
can be written in the form (\ref{2.9}) vanish. Consequently, the
complete contribution of the fermions to the effective action at zero
temperature is given by

\begin{equation}
\Gamma\,[A]=\frac{N_f}{2}\int dt\,A(t)=\frac{N_f}{2}\,a
\label{2.11}
\end{equation}

\bigskip\noindent
This is indeed the correct result.

\vspace{1cm}
\section {Various relations}
\renewcommand{\theequation}{3.\arabic{equation}}
\setcounter{equation}{0}

\bigskip
At finite temperature, the fermion propagator is given in the real
time formalism \cite{Das4} by  

\begin{equation}
S_F (p)=\frac{i}{p-m+i\epsilon}
-2\pi\,n_F(m)\,\delta(p-m)
\label{3.1}
\end{equation}

\bigskip\noindent
where

\begin{equation}
n_F(m)=\frac{1}{e^{\beta m}+1}
\label{3.2}
\end{equation}

\bigskip\noindent
is the Fermi statistical factor.

\medskip
The derivative expansion, as we have seen, involves a product of
propagators and, consequently, at finite temperature it would involve
evaluating the integral of the product of a number of zero
temperature propagators and a number of delta functions. These are
extremely singular integrals and so, we need to define them
carefully. In this section, we do precisely this which would help us
in the analysis of the derivative expansion at finite temperature.

\medskip
Let us denote a fermion loop (single flavor) with $N+1$ external
photon lines with all possible insertions of the photon momenta by

\begin{eqnarray}
i\,\tilde I_{N+1}&=&-\,\frac{(-\,i)^{N+1}}{(N+1)!}
\int\frac{dp}{2\pi}\sum_{perm}S_F(p)S_F(p+k_1)
S_F(p+k_1+k_2)
\nonumber\\
&&\phantom{-\,\frac{(-\,i)^{N+1}}{(N+1)!}
\int\frac{dp}{2\pi}\sum_{perm}S_F(p)}
\cdots S_F(p+k_1+\cdots+k_N)
\label{3.3}
\end{eqnarray}

\bigskip\noindent
Here ``{\it perm}" stands for all possible permutations of the photon
lines.  The $N=0$ term, or the one point function is the simplest and
has the form

\begin{eqnarray}
i\,\tilde I_1&=&i\int\frac{dp}{2\pi}\,
\biggl(\frac{i}{p-m+i\epsilon}-2\pi n_F(m)\delta(p-m)\biggr)
\nonumber\\
&=&\frac{i}{2}\,\Bigl(1-2n_F(m)\Bigr)
\label{3.4}
\end{eqnarray}

\bigskip\noindent
Any higher point function, on the other hand, will be a polynomial in
$n_F(m)$ starting with the linear term (no constants). Each of these
can be individually examined and since the power of $n_F(m)$
corresponds to the number of delta functions, we would have the
necessary relations.

\medskip
It is now straightforward to explicitly obtain the linear term in the
$N+1$ pt. function at lower orders.

\begin{eqnarray}
i\,\tilde I_2^{(1)}&=&-\,\frac{i}{2!}\,
\Bigl(\frac{1}{k+i\epsilon}-\frac{1}{k-i\epsilon}\Bigr)\,n_F(m)
\nonumber\\
&=&-\,\frac{i}{2!}\,\Bigl(-2i\pi\,\delta(k)\Bigr)\,n_F(m)
\nonumber\\
&=&-\,\frac{1}{2}\,n_F(m)\,\Bigl(2\pi\delta(k)\Bigr)
\nonumber\\
i\,\tilde I_3^{(1)}&=&-\,\frac{i}{3!}\,
\Bigl(\frac{1}{k_1+i\epsilon}\,\frac{1}{k_2+i\epsilon}
-\frac{1}{k_1+i\epsilon}\,\frac{1}{k_2-i\epsilon}
\nonumber\\
&&\phantom{-\,\frac{i}{3!}\,\,}
-\frac{1}{k_1-i\epsilon}\,\frac{1}{k_2+i\epsilon}
+\frac{1}{k_1-i\epsilon}\,\frac{1}{k_2-i\epsilon}\Bigr)\,n_F(m)
\nonumber\\
i\,\tilde I_4^{(1)}&=&-\,\frac{i}{4!}\,
\biggl[\frac{1}{(k_1+i\epsilon)(k_2+i\epsilon)(k_3+i\epsilon)}
\nonumber\\
&&\phantom{-\,\frac{i}{3!}\,\,}
-\frac{1}{k_1-i\epsilon}\,\frac{1}{(k_2+i\epsilon)(k_3+i\epsilon)}
\nonumber\\
&&\phantom{-\,\frac{i}{3!}\,\,}
-\frac{1}{k_2-i\epsilon}\,\frac{1}{(k_3+i\epsilon)(k_1+i\epsilon)}
\nonumber\\
&&\phantom{-\,\frac{i}{3!}\,\,}
-\frac{1}{k_3-i\epsilon}\,\frac{1}{(k_1+i\epsilon)(k_2+i\epsilon)}
\nonumber\\
&&\phantom{-\,\frac{i}{3!}\,\,}
+\frac{1}{(k_1-i\epsilon)(k_2-i\epsilon)}\,\frac{1}{k_3+i\epsilon}
\nonumber\\
&&\phantom{-\,\frac{i}{3!}\,\,}
+\frac{1}{(k_3-i\epsilon)(k_1-i\epsilon)}\,\frac{1}{k_2+i\epsilon}
\nonumber\\
&&\phantom{-\,\frac{i}{3!}\,\,}
+\frac{1}{(k_2-i\epsilon)(k_3-i\epsilon)}\,\frac{1}{k_1+i\epsilon}
\nonumber\\
&&\phantom{-\,\frac{i}{3!}\,\,}
-\frac{1}{(k_1-i\epsilon)(k_2-i\epsilon)(k_3-i\epsilon)}
\biggr]\,n_F(m)
\label{3.5}
\end{eqnarray}

\bigskip\noindent
and so on. The general pattern of terms for the linear term is now
clear. Let

\begin{equation}
J_n(k_1,k_2,\dots,k_n)=\frac{1}
{(k_1+i\epsilon)(k_2+i\epsilon)\cdots(k_n+i\epsilon)}
\label{3.6}
\end{equation}

\bigskip\noindent
Then, we can write

\begin{eqnarray}
i\,\tilde I^{(1)}_{N+1}&=&-\,\frac{i}{(N+1)!}\,
\biggl[J_N(k_1,k_2,\dots,k_N)
\nonumber\\
&&-\frac{1}{k_1-i\epsilon}\,J_{N-1}(k_2,k_3,\dots,k_N)-\cdots
-\frac{1}{k_N-i\epsilon}\,J_{N-1}(k_1,k_2,\dots,k_{N-1})
\nonumber\\
&&+\frac{1}{(k_1-i\epsilon)(k_2-i\epsilon)}\,
J_{N-2}(k_3,\dots,k_N)+\cdots
\nonumber\\
&&+\frac{1}{(k_{N-1}-i\epsilon)(k_N-i\epsilon)}\,
J_{N-2}(k_1,\dots,k_{N-2})-\cdots
\nonumber\\
&&+\cdots
\nonumber\\
&&+\cdots
\nonumber\\
&&+(-1)^N\,\frac{1}{(k_1-i\epsilon)(k_2-i\epsilon)
\cdots(k_N-i\epsilon)}\biggr]\,n_F(m)
\label{3.7}
\end{eqnarray}

\bigskip\noindent
From the definition in (\ref{3.6}), we see that we can combine terms
pairwise in (\ref{3.7}) to write

\begin{eqnarray}
i\,\tilde I_{N+1}^{(1)}&=&-\,\frac{i}{(N+1)!}\,
\Bigl(\frac{1}{k_1+i\epsilon}-\frac{1}{k_1-i\epsilon}\Bigr)
\biggl[J_{N-1}(k_2,k_3,\dots,k_N)
\nonumber\\
&&-\frac{1}{k_2-i\epsilon}\,J_{N-2}(k_3,\dots,k_N)-\cdots
-\frac{1}{k_N-i\epsilon}\,J_{N-1}(k_2,\dots,k_{N-1})
\nonumber\\
&&+\cdots
\nonumber\\
&&+(-1)^{N-1}\,\frac{1}{(k_2-i\epsilon)\cdots(k_N-i\epsilon)}
\biggr]\,n_F(m)
\nonumber\\
&=&\frac{1}{N+1}\,\Bigl(-\,2i\pi\,\delta(k_1)\Bigr)\,
i\tilde I_N^{(1)}
\label{3.8}
\end{eqnarray}

\bigskip\noindent
Upon iteration, this gives

\begin{eqnarray}
i\,\tilde I_{N+1}^{(1)}&=&\frac{(-i)^{N+1}}{(N+1)!}\,
\Bigl(2\pi\delta(k_1)\Bigr)\,\Bigl(2\pi\delta(k_2)\Bigr)
\cdots\Bigl(2\pi\delta(k_N)\Bigr)\,n_F(m)
\nonumber\\
&=&i\,I^{(1)}_{N+1}\,
\Bigl(2\pi\delta(k_1)\Bigr)\cdots\Bigl(2\pi\delta(k_N)\Bigr)
\label{3.9}
\end{eqnarray}

\bigskip
We can, similarly, analyze the terms quadratic and cubic in $n_F(m)$.
The details are not very illuminating and so we only give the final
results. 

\begin{eqnarray}
i\,\tilde I_{N+1}^{(2)}&=&i\,I^{(2)}_{N+1}\,
\Bigl(2\pi\delta(k_1)\Bigr)\cdots\Bigl(2\pi\delta(k_N)\Bigr)
\nonumber\\
i\,\tilde I_{N+1}^{(3)}&=&i\,I^{(3)}_{N+1}\,
\Bigl(2\pi\delta(k_1)\Bigr)\cdots\Bigl(2\pi\delta(k_N)\Bigr)
\label{3.10}
\end{eqnarray}

\bigskip\noindent
where

\begin{eqnarray}
i\,I_{N+1}^{(2)}&=&-\,\frac{(-i)^{N+1}}{(N+1)!}\,
\Bigl(\,^N\!C_1+^N\!C_2+\cdots+^N\!C_N\Bigr)\,\Bigl(n_F(m)\Bigr)^2
\nonumber\\
&=&-\,\frac{(-i)^{N+1}}{(N+1)!}\,
\Bigl(2^N-1\Bigr)\,\Bigl(n_F(m)\Bigr)^2
\nonumber\\
i\,I_{N+1}^{(3)}&=&\frac{(-i)^{N+1}}{(N+1)!}\,
\biggl[\,^N\!C_1\,\Bigl(\,^{N-1}\!C_1+^{N-1}\!C_2+^{N-1}\!C_3
+\cdots+^{N-1}\!C_{N-1}\Bigr)
\nonumber\\
&&\phantom{\frac{(-i)^{N+1}}{(N+1)!}\,}
+^N\!C_2\,\Bigl(\,^{N-2}\!C_1+^{N-2}\!C_2+^{N-2}\!C_3
+\cdots+^{N-2}\!C_{N-2}\Bigr)
\nonumber\\
&&\phantom{i\,I_{N+1}^{(3)}=\frac{(-i)^{N+1}}{(N+1)!}\,}
+\cdots
\nonumber\\
&&\phantom{i\,I_{N+1}^{(3)}=\frac{(-i)^{N+1}}{(N+1)!}\,}
+^N\!C_{N-1}\,\Bigl(\,^1C_1\Bigr)\biggr]\,\Bigl(n_F(m)\Bigr)^3
\nonumber\\
&=&\frac{(-i)^{N+1}}{(N+1)!}\Bigl(3^N-2\cdot2^N+1\Bigr)\,\Bigl(n_F(m)
\Bigr)^3
\label{3.11}
\end{eqnarray}

\bigskip\noindent
In this way, we can derive the general formula that (for $r\leq N+1$
and $r\geq1$)

\begin{equation}
i\,\tilde I_{N+1}^{(r)}=i\,I_{N+1}^{(r)}\,
\Bigl(2\pi\,\delta(k_1)\Bigr)\cdots
\Bigl(2\pi\,\delta(k_N)\Bigr)
\label{3.12}
\end{equation}

\bigskip\noindent 
where

\begin{eqnarray}
i\,I_{N+1}^{(r)}&=&(-1)^{r+1}\,\frac{(-i)^{N+1}}{(N+1)!}\,
\Bigl(n_F(m)\Bigr)^r
\sum_{s=0}^{r-1}(-1)^s\,^{r-1}\!C_s\,(r-s)^N
\nonumber\\
&=&(-1)^{r+1}\,\frac{(-i)^{N+1}}{(N+1)!}\,
\Bigl(n_F(m)\Bigr)^r\,b^{(r)}_{N+1}
\label{3.13}
\end{eqnarray}

\bigskip\noindent
so that the $N+1$ pt. function can be written as $(\tilde I_1=I_1)$

\begin{eqnarray}
i\,\tilde I_{N+1}&=&\Bigl(2\pi\,\delta(k_1)\Bigr)\cdots
\Bigl(2\pi\,\delta(k_N)\Bigr)\,\sum_{r=1}^{N+1}i\,I_{N+1}^{(r)}
\nonumber\\
&=&\Bigl(2\pi\,\delta(k_1)\Bigr)\cdots
\Bigl(2\pi\,\delta(k_N)\Bigr)\,i\,I_{N+1}
\label{3.14}
\end{eqnarray}

\bigskip
The results in (\ref{3.12})--(\ref{3.14}) are interesting from
various points of view. First, from the explicit form of the result
in (\ref{3.14}), we see that 

\begin{equation}
k_i\,\tilde I_{N+1}(k_1,\dots,k_N)=0
\hspace{1cm}{\rm for}\,\,i=1,2,\dots,N
\label{3.14a}
\end{equation}

\bigskip\noindent
verifying explicitly that the amplitudes satisfy the Ward identities
of the theory. In this sense, it is complementary to the analysis in
ref. \cite{Das1} where the Ward identities were used to carry out the
calculations. Second, from the properties of the binomial
coefficients, it is straightforward to see that

\begin{equation}
b_{N+1}^{(r)}=(r-1)\,b_N^{(r-1)}+r\,b_N^{(r)}
\label{3.15}
\end{equation}

\bigskip\noindent
which leads to the relation that

\begin{equation}
\frac{\partial I_N}{\partial m}
=-\,i\beta(N+1)\,I_{N+1}
\label{3.16}
\end{equation}

\bigskip\noindent
which is the recursion relation derived in ref. \cite{Das1} following
from the Ward identity and we can identify $\tilde I_N$ with the 
vertex
functions obtained in ref. \cite{Das1} up to a flavor factor.
Finally, if we set all the external momenta to zero, then from
(\ref{3.13}), we obtain (for $r\leq N+1$ and $r\geq1$)

\begin{equation}
\int\frac{dp}{2\pi}\,\frac{1}{(p-m+i\epsilon)^{N+1-r}}\,
\delta^r(p-m)=(-i)^{N+1-r}\,b_{N+1}^{(r)}\,
\Bigl(2\pi\,\delta(0)\Bigr)^N
\label{3.17}
\end{equation}

\bigskip\noindent
This defines the singular integrals consistent with the Ward
identity. With these relations, we are now ready to analyze the
derivative expansion at finite temperature.

\vspace{1cm}
\section{Derivative expansion at T$\neq$0}
\renewcommand{\theequation}{4.\arabic{equation}}
\setcounter{equation}{0}

\bigskip
Going back to the definition of the derivative expansion in
(\ref{2.4}), we note that the first term in the expansion is again
quite simple at finite temperature (as is derived in (\ref{3.4})
also). 

\begin{eqnarray}
\Gamma_1[A]&=&-\,i\,N_f\,{\rm Tr}\,iS_F(p)\,A(t)
\nonumber\\
&=&-\,i\,N_f\int\frac{dp}{2\pi}\,
\Bigl(-\,\frac{1}{p-m+i\epsilon}
-2i\pi\,n_F(m)\delta(p-m)\Bigr)\,\int dt\,A(t)
\nonumber\\
&=&\frac{N_f}{2}\,\Bigl(1-2n_F(m)\Bigr)\,\int dt\,A(t)
\nonumber\\
&=&\frac{N_f}{2}\,\Bigl(1-2n_F(m)\Bigr)\,a
\nonumber\\
&=&N_f\,I_1\,a
\label{4.1}
\end{eqnarray}

\bigskip\noindent
It is at the second and higher orders that we run into the conceptual
question of how a single time integral can become a nonextensive
term. Therefore, let us analyze the second order term in some detail.

\begin{eqnarray}
\Gamma_2[A]&=&i\,\frac{N_f}{2}\,{\rm Tr}\,i\,S_F(p)\,A\,
\bigl(iS_F(p)\bigr)\,A
\nonumber\\
&=&-\,i\,\frac{N_f}{2}\,{\rm Tr}\,S_F(p)\,
S_F(p+i\partial_1)\,A_1A_2
\label{4.2}
\end{eqnarray}

\bigskip\noindent
Here we have used the commutation relations in (\ref{2.5}) to move
all the momentum dependent factors to the left and have labeled the
fields $A$ merely to indicate that the derivative acts only on the
first of the two fields. The trace can now be taken and the momentum
integral can be done using (\ref{3.14}) which gives

\begin{equation}
\Gamma_2[A]=N_f\,I_2\int dt\,2\pi\,
\Bigl(\delta(\partial_t)A\Bigr)\,A
\label{4.3}
\end{equation}

\bigskip\noindent
This is a single integral, but the integrand is quite unusual. It is
clear from the properties of delta functions that
$(\delta(\partial_t)A)$ cannot depend on $t$. In fact, it is much
simpler to analyze this in the Fourier transformed space where it
becomes 

\begin{eqnarray}
\Gamma_2[A]&=&N_f\,I_2\int\frac{dp}{2\pi}\,
2\pi\,\delta(k)\tilde A(k)\,\tilde A(-k)
\nonumber\\
&=&N_f\,I_2\,\tilde A(0)\,\tilde A(0)
\nonumber\\
&=&N_f\,I_2\,\Bigl(\int dt\,A(t)\Bigr)^2
\nonumber\\
&=&N_f\,I_2\,a^2
\label{4.4}
\end{eqnarray}

\bigskip
This is quite interesting and clarifies how a single time integral
becomes a nonextensive term in the effective action. This also brings
out another interesting point. Namely, had we expanded the second
propagator in (\ref{4.2}) in powers of the derivatives, as the
philosophy of the derivative expansion would dictate, we would have
obtained an infinite series of local terms. Each momentum integral,
in this case, can be done using (\ref{3.16}). If we sum this series,
of course, we would obtain the nonextensive term in (\ref{4.4}).
However, if we were to truncate the powers of the derivative at some
order, as would become the case in any realistic, nontrivial model,
we would miss completely the nonextensive nature of the effective
action. The nonextensive structure is, of course, quite crucial in
restoring the large gauge invariance. So, it would appear that the
derivative expansion is not an ideal setting to discuss the question
of large gauge invariance.

\medskip
To complete this calculation, then, at the $(N+1)$th order, if we
commute all the propagators to the left and do the momentum integral
using (\ref{3.13}), we would obtain

\begin{eqnarray}
\Gamma_{N+1}&=&N_f\,I_{N+1}\int dt\,
\Bigl(2\pi\delta(\partial_t)A\Bigr)\cdots
\Bigl(2\pi\delta(\partial_t)A\Bigr)\,A
\nonumber\\
&=&N_f\,I_{N+1}\,\Bigl(\int dt\,A(t)\Bigr)^{N+1}
\nonumber\\
&=&N_f\,I_{N+1}\,a^{N+1}
\label{4.5}
\end{eqnarray}

\bigskip\noindent
(The correct combinatoric factor arises from appropriate
symmetrization.) This is exactly the result obtained in ref.
\cite{Das1} (remember that $N_f\,I_f\rightarrow I_N$ of ref.
\cite{Das1}) and hence the series can again be summed to give

\begin{eqnarray}
\Gamma[A]&=&\sum_{N=1}^\infty\Gamma_N
\nonumber\\
&=&-\,i\,N_f\,\ln\,\bigl(\cos\frac{a}{2}+i\,
\tanh\frac{\beta m}{2}\,\sin\frac{a}{2}\bigr)
\label{4.6}
\end{eqnarray}

\bigskip\noindent 
which is the exact result \cite{Du}.

\vspace{1cm}
\section{Conclusion}

\bigskip
In this paper, we have solved the $0+1$ dimensional Chern-Simons
theory within the framework of the derivative expansion. We have
derived various interesting relations that will be useful in the
study of finite temperature field theory. We have also explicitly
clarified how a single time integral gives rise to nonextensive terms
in the effective action. It appears from this analysis that the
derivative expansion is not an ideal setting to examine the question
of large gauge invariance in higher dimensions.

\vspace{1cm}
\noindent {\bf Acknowledgment:} A.D. would like to thank the members
of the Department of Theoretical Physics in UFRJ for hospitality
where this work was done. J.B.-N. is supported in part by
Conselho Nacional de Desenvolvimento Cient\'{\i}fico e Tecnol\'ogico
- CNPq, Financiadora de Estudos e Projetos - FINEP, and
Funda\c{c}\~ao Universit\'aria Jos\'e Bonif\'acio - FUJB (Brazilian
Research Agencies). A.D. is supported in part by US DOE Grant No.
DE-FG-02-91ER40685, NSF-INT-9602559 and a Fulbright grant.


\vspace{1cm}
\begin{thebibliography}{30}
\bibitem{Deser1} S. Deser, R. Jackiw and S. Templeton, Phys. Rev.
Lett. 48 (1982) 975; Ann. Phys. 140 (1982) 372.
\bibitem{Re} A. Redlich, Phys. Rev. Lett. 52 (1984) 18; Phys.
Rev. D29 (1984) 2366.
\bibitem{Pi} R.D. Pisarski, Phys. Rev. D35 (1987) 664.
\bibitem{Ishi} K.T. Ishikawa and A. Matsuyama, Nucl. Phys. B280 
(1987)
523. 
\bibitem{Babu} K.S. Babu, A. Das and P. Panigrahi, Phys. Rev. D36
(1987) 3725.
\bibitem{Bra} N. Brali\'c, C. Fosco and F. Schaposnik, Phys. Lett.
B383 (1996) 199; D. Cabra, E. Fradkin, G. Rossini and F. Schaposnik,
Phys. Lett. B383 (1996) 434.
\bibitem{Du} G. Dunne, K. Lee and C. Lu, Phys. Rev. Lett. 78 (1997)
3434. 
\bibitem{Das1} A. Das and G. Dunne, hep-th/9712144, to appear in 
Phys.
Rev. D.
\bibitem{Deser2} S. Deser, L. Griguolo and D. Seminara, Phys. Rev. 
Lett.
79 (1997) 1976; S. Deser, L. Griguolo and D. Seminara,
hep-th/9712066. 
\bibitem{Fos} C. Fosco, G. Rossini and F. Schaposnik, Phys. Rev. 
Lett.
79 (1997) 1980; Phys. Rev. D56 (1997) 6547.
\bibitem{Ait1} I. Aitchison and C. Fosco, hep-th/9709035.
\bibitem{Gon} R. Gonz\'ales-Felipe, hep-th/9709079.
\bibitem{Ait2} I.Aitchison and C. Fraser, Phys. Lett. 146B (1984) 63.
\bibitem{Das2} A. Das and A. Karev, Phys. Rev. D36 (1987) 623; Phys.
Rev. D36 (1987) 2591.
\bibitem{Das3} A. Das and M. Hott, Phys. Rev. D50 (1994) 6655.
\bibitem{Das4} A. Das, {\it Finite temperature field theory}, World
Scientific (1997)
\end{thebibliography}
\end{document}